\documentclass[twocolumn,citeautoscript,pra,superscriptaddress,amsmath,amssymb,8pt]{revtex4-1}
\usepackage{graphicx}
\usepackage{gensymb}
\usepackage{color}
\usepackage[dvipsnames]{xcolor}
\usepackage{float}
\usepackage{upgreek}
\usepackage{amsmath,amssymb}

\graphicspath{{C:/Users/Alex/Desktop/plots/}}

\begin{document}

\title{Hyperpolarisation of external nuclear spins using nitrogen-vacancy centre ensembles}

\author{A. J. Healey}
\affiliation{School of Physics, University of Melbourne, VIC 3010, Australia}
\affiliation{Centre for Quantum Computation and Communication Technology, School of Physics, University of Melbourne, VIC 3010, Australia}	

\author{L. T. Hall}
\affiliation{School of Physics, University of Melbourne, VIC 3010, Australia}

\author{G. A. L. White}
\affiliation{School of Physics, University of Melbourne, VIC 3010, Australia}

\author{T. Teraji}
\affiliation{National Institute for Materials Science, Tsukuba, Ibaraki 305-0044, Japan}

\author{M.-A. Sani}
\affiliation{School of Chemistry, Bio21 Institute, University of Melbourne, VIC 3010, Australia}

\author{F. Separovic}
\affiliation{School of Chemistry, Bio21 Institute, University of Melbourne, VIC 3010, Australia}

\author{J.-P. Tetienne}
\email{jtetienne@unimelb.edu.au}
\affiliation{School of Physics, University of Melbourne, VIC 3010, Australia}
\affiliation{Centre for Quantum Computation and Communication Technology, School of Physics, University of Melbourne, VIC 3010, Australia}	

\author{L. C. L. Hollenberg}
\email{lloydch@unimelb.edu.au}
\affiliation{School of Physics, University of Melbourne, VIC 3010, Australia}
\affiliation{Centre for Quantum Computation and Communication Technology, School of Physics, University of Melbourne, VIC 3010, Australia}

\begin{abstract}
The nitrogen-vacancy (NV) centre in diamond has emerged as a candidate to non-invasively hyperpolarise nuclear spins in molecular systems to improve the sensitivity of nuclear magnetic resonance (NMR) experiments. Several promising proof of principle experiments have demonstrated small-scale polarisation transfer from single NVs to hydrogen spins outside the diamond. However, the scaling up of these results to the use of a dense NV ensemble, which is a necessary prerequisite for achieving realistic NMR sensitivity enhancement, has not yet been demonstrated. In this work, we present evidence for a polarising interaction between a shallow NV ensemble and external nuclear targets over a micrometre scale, and characterise the challenges in achieving useful polarisation enhancement. In the most favourable example of the interaction with hydrogen in a solid state target, a maximum polarisation transfer rate of $\approx 7500$ spins per second per NV is measured, averaged over an area containing order $10^6$ NVs. Reduced levels of polarisation efficiency are found for liquid state targets, where molecular diffusion limits the transfer. Through analysis via a theoretical model, we find that our results suggest implementation of this technique for NMR sensitivity enhancement is feasible following realistic diamond material improvements. 
\end{abstract}

\maketitle 

\section{Introduction}
Nuclear magnetic resonance (NMR) underpins a variety of techniques that find use across the fields of physics, chemistry, and the life sciences. The sensitivity of an NMR measurement is proportional to the degree of nuclear spin polarisation in the target to be analysed, which is very low under typical thermal conditions ($P_{\text{th}}\approx 10^{-5}$ at room temperature and a magnetic field of 3~T). Consequently, a number of approaches have been developed to achieve levels of polarisation well in excess of thermal levels (``hyperpolarisation") in nuclear spin ensembles. To date, dynamic nuclear polarisation (DNP) \cite{Abragam1978,Wind1985,Hall1997,Tateishi2014} is the most widely implemented method of achieving hyperpolarisation, with para-hydrogen induced polarisation (PHIP) \cite{Natterer1997, Hovener2018} and optical pumping \cite{Walker1997,Navon1996} based methods also finding success. These methods, however, are far from being problem-free in their application. Each suffers from limitations, such as in their target specificity (PHIP, optical pumping), or in the technically challenging conditions required for optimal operation (e.g. cryogenic temperatures, large magnetic fields, and strong microwave driving for DNP). 

For these reasons, it has been proposed that the negatively charged nitrogen-vacancy (NV) centre defect in diamond \cite{Doherty2013} may be a suitable candidate to address some of the shortcomings in other techniques \cite{Broadway2018a,Fernandez-Acebal2018,Shagieva2018}. The NV centre's electron spin is efficiently polarised ($\approx 80 \%$) on a microsecond time scale and is easily manipulated at room temperature. Promising proof of principle experiments have demonstrated polarisation transfer from NV centres to nuclear targets via the magnetic dipole-dipole interaction using a variety of experimental protocols \cite{Broadway2018a,Fernandez-Acebal2018,Shagieva2018,London2013,King2010,Pagliero2018,Ajoy2018,Henshaw2019,Schwartz2018,Lang2019}. This transfer is non-invasive to the target and general, avoiding the target specificity limitations of other techniques. Further, the potential for room temperature and low-field operation raises the prospect of achieving hyperpolarisation with a reduced technical overhead. However, work so far has been limited to either internal $^{13}$C \cite{London2013,King2010,Pagliero2018,Ajoy2018,Henshaw2019,Schwartz2018,Lang2019} or small-scale external polarisation with single NV centres \cite{Broadway2018a,Fernandez-Acebal2018,Shagieva2018}, which are naturally limited in their scope towards achieving bulk hyperpolarisation over a sample volume useful for NMR. Recent theoretical work in Ref. \cite{Tetienne2021} showed that bulk NMR enhancement (as well as for NV-based, micron-scale NMR \cite{Glenn2018,Smits2019}) is possible using dense NV ensembles with sufficiently good quantum coherence properties. However, successful experimental demonstration of polarisation transfer from an NV ensemble to an external nuclear target has not yet been achieved.

Although NV ensembles essentially act as many independent NVs for the purpose of hyperpolarisation \cite{Tetienne2021}, extending the previously reported single NV results to the case of a dense ensemble is not trivial. As the distance between NVs and nuclei external to the diamond will be several nanometres, the dipole-dipole coupling that governs the polarisation transfer is weak, and thus the NV coherence time $T_2$ (which is scheme dependent) emerges as the limiting variable \cite{Hall2020}. In increasing the density of NV centres within the diamond sample, the magnetic noise is increased proportionally, degrading these quantum properties. The dominant noise source in a dense ensemble is the substitutional nitrogen bath \cite{Bauch} which, with current standard sample production techniques, is at minimum ten times more abundant than the NV density \cite{Pezzagna2010}. In addition, the necessity of placing the NVs near the diamond surface brings another significant contribution to NV decoherence: a range of fast-fluctuating surface defects that render near-surface NV properties much worse than those in the bulk and result in band bending that reduces the charge stability of the negative NV charge state within a few nanometres of the surface \cite{Stacey2019,Rosskopf2014,Myers2014,Romach2015,Bluvstein2019}. 

Further, while it may be possible to find a single NV located within an anomalously calm spin environment that exhibits unusually long coherence times, or that is situated anomalously close to the surface and thus couples more strongly to external spins, in dealing with an NV ensemble the resulting dynamics will be governed by average NV properties \cite{Tetienne2018}. It is not immediately clear, therefore, whether previous successful single NV results are easily extendible to higher NV densities. 

In this work, we present an experimental study into the challenges associated with the scaling up of previous single NV results \cite{Broadway2018a,Fernandez-Acebal2018,Shagieva2018} to shallow NV ensembles, with the goal of demonstrating polarisation transfer compatible with the vision of Ref. \cite{Tetienne2021}. We investigate the polarisation dynamics over a range of parameters using a robust, pulse-based scheme (PulsePol \cite{Schwartz2018}). Starting with an ideal room temperature scenario, we polarise a solid target (biphenyl) and analyse our results using a theoretical model. In doing so, we experimentally determine an upper bound on the polarisation (or cooling) rate in the current ``best case" scenario and discuss the implications of this. We then probe the extent to which the addition of molecular diffusion compromises the polarisation transfer by repeating experiments using fluid targets of varying viscosities. We conclude with a discussion of the results and an assessment of the prospects for future work. 

\section{Results}

\subsection{Polarising a solid target}
\label{sec: biphenyl}
\begin{figure}
\includegraphics[width=0.45\textwidth]{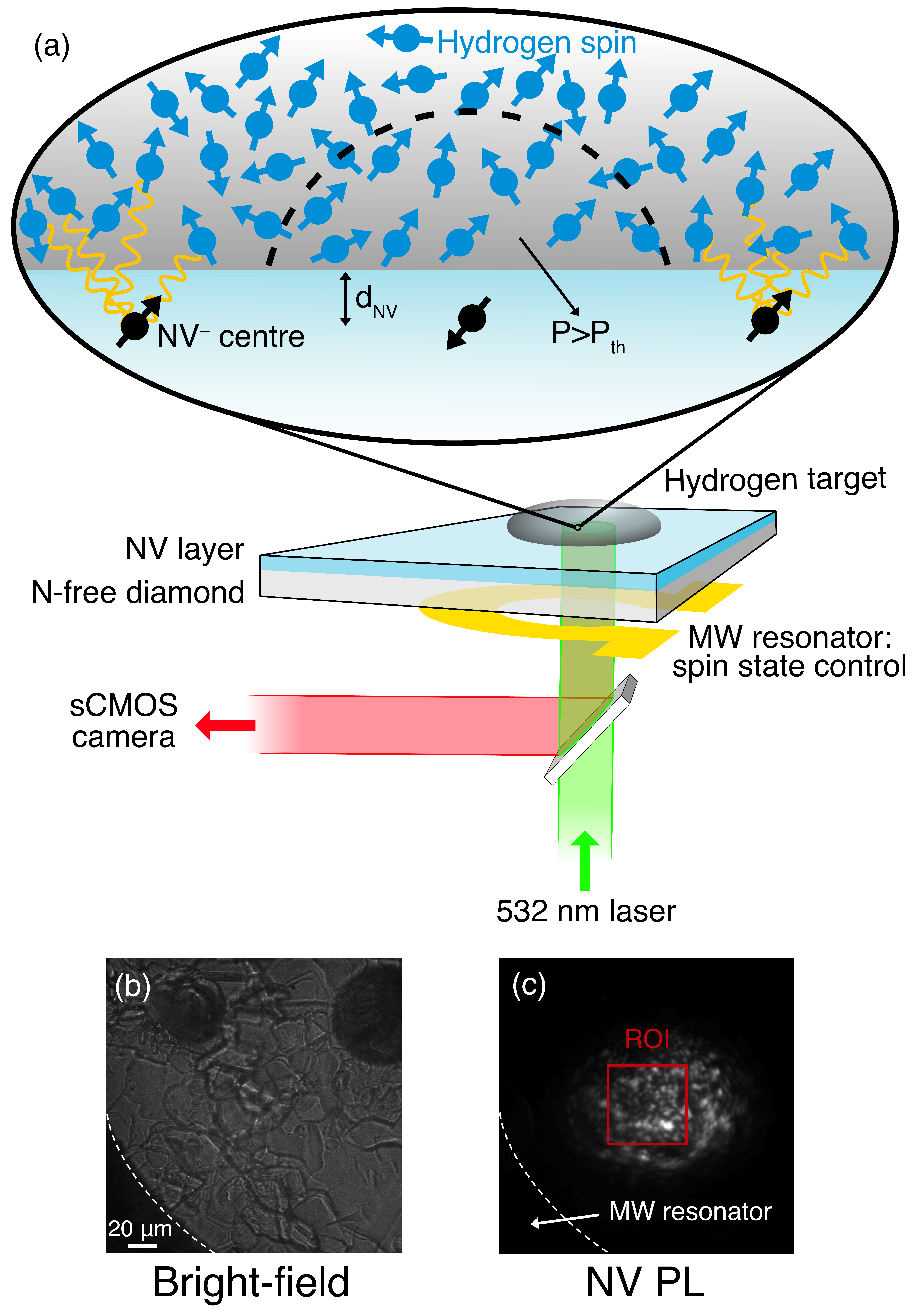}
\caption{\textbf{a)} Schematic of experimental setup, depicting widefield NV microscope and polarisation transfer via dipolar coupling between an ensemble of NVs at a depth $d_{NV}$ below the diamond surface and a number of nearby nuclear spins. Polarisation $P$ greater than the thermal background ($P_{\text{th}}$) is built up locally to an NV (dashed semicircle) as its spin polarisation (quantised along external bias field $B$) is donated to the nuclear bath, before spin diffusion spreads this polarisation into the bulk target. \textbf{b)} Bright-field image of a 200~$\times$~200~$\mu$m region, showing biphenyl crystals on the diamond surface. \textbf{c)} NV photolumiescence (PL) image of the same region in (b), with the 50~$\times$~50~$\mu$m region of interest (ROI) used for this work marked by the red square.}
\label{schematic}
\end{figure}
We first investigate the interaction between our shallow NV ensemble and hydrogen spins within a solid target. All experiments were carried out over a 50$\times$50~$\mu$m field of view (FOV) using a widefield NV microscope. Elements of this setup are depicted schematically in Fig. \ref{schematic} (see Appendix \ref{expt detail} for details), with the FOV denoted by the red square in the NV photoluminescence (PL) image, chosen to coincide with a region of relative laser intensity and microwave driving uniformity. Biphenyl crystals were formed on the surface of the diamond and encapsulated with epoxy to prevent subsequent sublimation (details in Appendix \ref{characterisation}). As our experiment addresses a large number of NVs simultaneously and a $\sim 100~\mu$m$^3$ target volume, it can be considered as a ``toy model" for a realistic, NMR-relevant implementation of the technique. Achieving polarisation over this scale presents two principal challenges: the technical, experimental challenge of successfully addressing a large number of NVs over a useful field of view, and the material problem of reduced coherence times for high density NV ensembles. The PulsePol pulse sequence has been identified as the best currently accessible approach in addressing these concerns due to its robustness to realistic experimental errors (see Appendix \ref{protocol comparison}) \cite{Schwartz2018}. Its action as a decoupling sequence is also beneficial in addressing our ensembles as they are subject to broadband noise from the nitrogen spin bath and surface defects. The ensembles were created using a 2.5~keV $^{15}$N implant with a fluence of $1$-$2\times10^{13}$~cm$^{-2}$ and a 1100$\degree$C ramped anneal, leading to NVs lying within 10~nm of the surface with an areal density $\sigma_{NV} \approx 1500~\mu$m$^{-2}$ (see details in Appendix \ref{characterisation}). Coherence times in excess of 10~$\mu$s were achieved using PulsePol which is a considerable extension over low-order decoupling sequences in this regime \cite{Tetienne2018,Bauch}.

\begin{figure}
\centering
\includegraphics[width=0.35\textwidth]{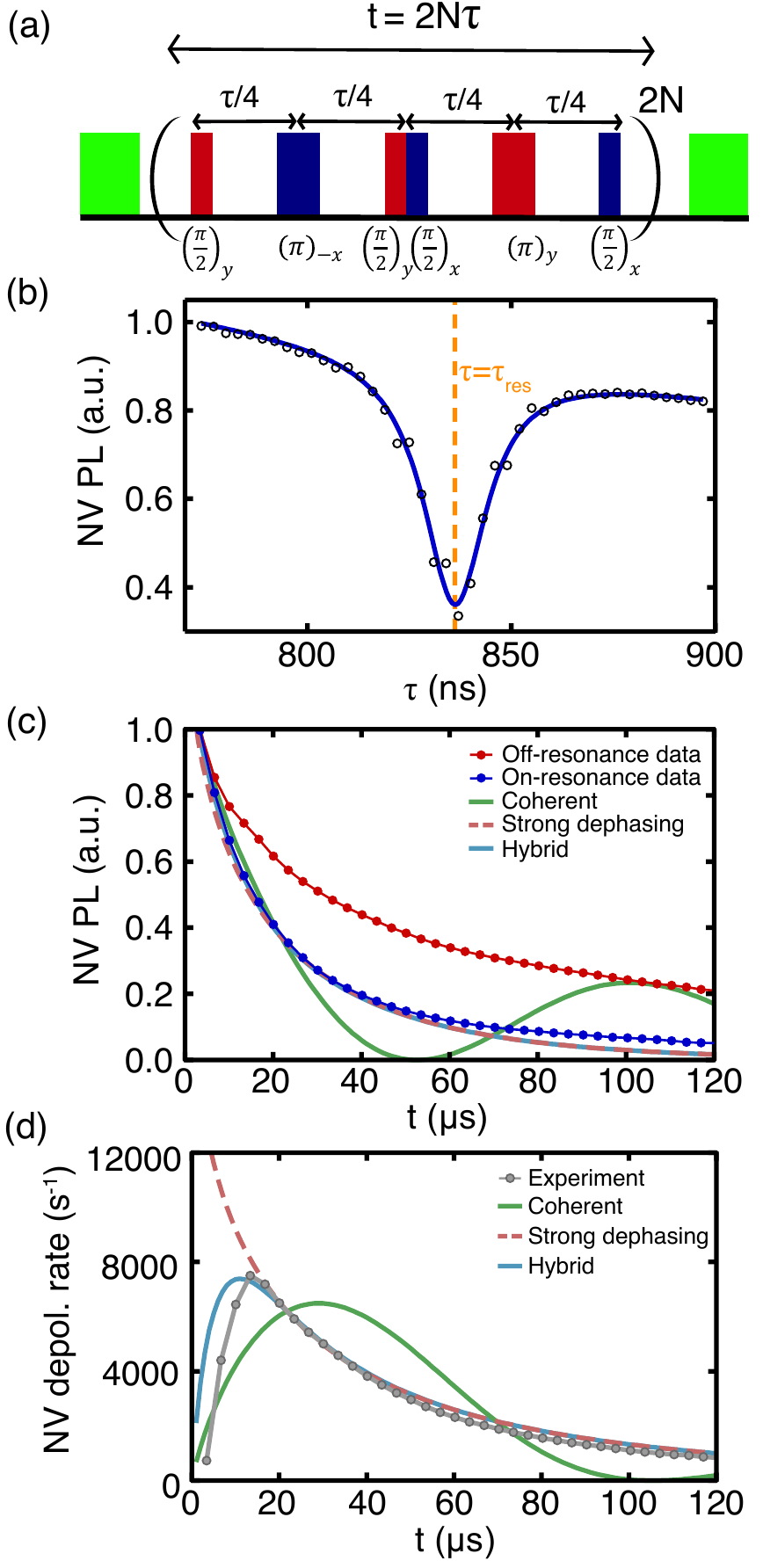}
\caption{\textbf{a)} Schematic of the PulsePol pulse sequence. \textbf{b)} $N=30$ PulsePol spectrum scanning $\tau$, with $T_{2,n}^*$-limited hydrogen resonance visible at $\tau = 3\tau_L/2$. The solid blue line is a fit to the data (black circles) using a Lorentzian lineshape. \textbf{c)} Off- (red) and on-resonance (dark blue) NV decay with PulsePol sequences of increasing length and total duration $t=2N\tau$. Solid lines give theoretical predictions based on numerically integrated values of $A_0$, and $\Gamma_2^{\text{tot}}$ for the coherent (green), strong-dephasing (light red), and hybrid (light blue) cases. \textbf{d)} Experimentally determined NV depolarisation rate (grey circles) and comparison to theory (solid curves).}
\label{fig: biphenyl}
\end{figure}

The PulsePol pulse sequence, depicted in Fig. \ref{fig: biphenyl}(a), sets an average Hamiltonian that approximates a flip-flop Hamiltonian between the NV and a target nuclear spin species when a resonant condition $\tau = n\tau_L/2$ for odd integer $n$ is met, where $\tau$ is defined as the duration of the unit sequence, repeated $2N$ times, and $\tau_L$ is the target Larmor period.

Fig. \ref{fig: biphenyl}(b) shows a typical PulsePol spectrum obtained by sweeping $\tau$ for fixed $N=30$. The y-axis is the normalised NV PL averaged over the full FOV, which is a proxy for the population of the $m_s=0$ NV spin state, referred to as NV spin polarisation in what follows. The dip at $\tau\approx 840$~ns is the $n=3$ hydrogen resonance, which produces the strongest interaction \cite{Schwartz2018} and will be the focus of this work. When the resonant condition is met, a large drop in NV spin polarisation is observed, which is inferred to have been symmetrically donated to the target hydrogen bath by virtue of the PulsePol sequence design \cite{Schwartz2018}. To gain an insight into the dynamics of the apparent polarisation transfer, we examine the decay of NV spin polarisation versus total interaction time $t=2N\tau$ by increasing $N$ for fixed $\tau$, and compare the resonant ($\tau = \tau_{\text{res}}$) case with the off-resonant case. Both sets of data are shown in Fig. \ref{fig: biphenyl}(c), where the red points are estimates of the off-resonance decay at $\tau=\tau_{\text{res}}$ made by averaging data obtained at $\tau=\tau_{\text{res}}\pm 60$~ns, and the blue points show the resonant decay. Following the treatment of Refs. \cite{Hall2016,Broadway2018,Hall2020},  we represent the resonant decay curve as a product of the background NV decoherence (which is independent of any polarisation dynamics) and the resonant contribution,

\begin{equation}
P_{NV}(t) = P_{\text{off}}(t)P_{\text{res}}(t),
\end{equation}
where $P_{NV}$ is the population of the NV m$_s=0$ spin state (assumed to be perfectly initialised at $t=0$).

For the present experiment, we find that the off-resonance decay is well fit by $P_{\text{off}}(t) = \exp(-(t /T_2^{NV})^\beta)$ with $T_2^{NV}=22~\mu$s and $\beta = 0.42$. In analysing the shape of the resonant decay curve, it is useful to consider the extreme cases of the strong and negligible dephasing regimes. In both cases, we assume that there is a direct correspondence between extra coherence lost by the NV when the resonant condition is met and that resulting from the flip-flop interaction in the effective Hamiltonian. 

In the ideal case, there is coherent coupling between NV and the hydrogen spin bath, and the evolution of the system is governed by
\begin{equation}
P_{\text{res}}(t) = \cos^2\left(\frac{A_{0}t}{2}\right),
\label{eqn coherent}
\end{equation}
where $A_0 = \sqrt{\displaystyle\sum_{j}A_j^2}$ is the summed dipolar coupling between the NV and full spin bath, made up of $j$ hydrogen spins \cite{Broadway2018a,Hall2020}. In the absence of decoherence, the flip-flop time $\tau_0 = \pi /A_0$ corresponds to the full donation of the NV's spin polarisation to the target bath. 

Conversely, in the strong dephasing regime $\Gamma_2^{\text{tot}} \gg A_0$, where $\Gamma_2^{\text{tot}} = \Gamma_2^{NV} + \Gamma_2^{H}$ (with $\Gamma_2^{NV} = 1/T_2^{NV}$ and $\Gamma_2^H$ is the hydrogen spin dephasing rate) is the total dephasing rate of the system under the PulsePol sequence, incoherent polarisation transfer proceeds via the monotonic evolution \cite{Broadway2018a,Hall2020}
\begin{equation}
P_{\text{res}}(t) = \exp \left(-\frac{A_0^2}{\Gamma_2^{\text{tot}}}t\right).
\label{eqn strong dephasing}
\end{equation} 

With an average NV depth measured at $d_{NV} \approx 6$~nm (see Appendix \ref{characterisation} for details), we expect our experiment to fall primarily in the strong dephasing regime \cite{Hall2020}. 

Fig. \ref{fig: biphenyl}(c) shows the extremal cases as predicted by numerical integration of the quantities $A_0$ and $\Gamma_2^{\text{tot}}$ (which is affected by a nuclear dephasing gradient due to the presence of unpaired electron spins on the diamond surface \cite{Hall2020}) with $d_{NV} = 6$~nm. Good agreement is found with the strong dephasing case, indicating that Eqn. \ref{eqn strong dephasing} gives a good approximation to the polarisation dynamics in our experiment. It is possible, however, that there are some target spins that exist in or near the strong coupling ($A_0 > \Gamma_2^{\rm tot}$) regime, implying the existence of a small coherent component in our data. In order to capture this interchange between coherent and incoherent behaviour, as averaged over our measurement ensemble, we introduce the following phenomenological adjustment to Eqn. \ref{eqn strong dephasing}:

\begin{equation}
\label{eqn: combined}
\begin{split}
P_{NV}(t) & = \exp \left(-(\Gamma_2^{NV}t)^{\beta} \right) \left\{ \exp (- \Gamma_{\rm int} t) \cos ^2\left(\frac{A_0 t}{2} \right) \right.\\ &\left.+ \left(1-\exp (-\Gamma_{\rm int} t)\right)\exp\left(-\frac{A_0^2}{\Gamma_2^{\text{tot}}}t\right)\right\},
\end{split}
\end{equation}

where $\Gamma_{\rm int}$ is a parameter controlling this interchange. Setting $\Gamma_{\rm int} \approx \Gamma_2^{\text{tot}} \approx 100$~kHz makes only a subtle adjustment to the decay shape, however the influence of this factor becomes more apparent in Fig. \ref{fig: biphenyl}(d). Here, we plot the difference between the experimental off- and on-resonance data, normalised by the difference in PL given by ensembles initialised in the $m_s = 0$ and $m_s = -1$ spin-states (representing maximal NV coherence), and divided by $t$ at each point. This amounts to a direct experimental measurement of the rate of additional NV depolarisation due to the interaction with the hydrogen bath, which we infer to correspond directly with the hydrogen cooling rate, defining
\begin{equation}
\label{eqn: cooling rate}
u(t) = \frac{1}{t} \left( P_{\text{off}}(t) - P_{NV}(t)\right),
\end{equation} 
which is a generalisation of the definition in Refs. \cite{Tetienne2021,Hall2020}. Note that this equation is exact when the hydrogen bath polarisation is given by $P_H(t)=P_{\text{off}}(t) - P_{NV}(t)$ and an overestimate when there is additional resonant coherence loss that does not contribute to useful polarisation. The current experiment is unable to make the distinction and so the cooling rates quoted in this work represent upper limits of the true values.

The experimental data peaks at a value of $u\approx 7500$~s$^{-1}$ for $N=8$, corresponding to an interaction time $t\approx 13$~$\mu$s. The strong dephasing approximation vastly overestimates the NV depolarisation rate for small times because it sees coherence decrease exponentially while coherent transfer is sinusoidal ($\sin^2 x \approx x^2$ for small $x$), while the fully coherent case predicts an accurate maximum depolarisation rate but for a much longer sequence than the actual $N=8$. The modification of Eqn. \ref{eqn: combined} is required to give qualitative agreement with the data, producing a peak positioned according to the value of $\Gamma_{\rm int}$. The good agreement we find with a model that contains only experimentally determined parameters and one free parameter $\Gamma_{\rm int}$ suggests that the assumption that NV depolarisation corresponds to hydrogen bath polarisation is valid. Additionally, the close correspondence between the magnitude of the maximum experimental cooling rate and that predicted in the coherent case (accounting for background NV decay) supports the use of the coherent model for general predictions as in Ref. \cite{Tetienne2021}. 

Under continuous application of this optimal sequence, polarisation can be built up within the target bath according to the differential equation:
\begin{equation}
\label{eqn: diffeq}
\begin{split}
\frac{\partial P(\textbf{R},T)}{\partial T} = &u(\textbf{R})\left[1-P(\textbf{R},T)\right] - \Gamma_{1,n}P(\textbf{R},T)\\ &+ D_n\nabla^2P(\textbf{R},T),
\end{split}
\end{equation}
where $\Gamma_{1,n}$ is the nuclear relaxation rate, $D_n$ is the nuclear diffusion constant, and $u(\textbf{R})$ is the position-dependent cooling rate, integrating to the maximum experimentally determined NV depolarisation rate \cite{Tetienne2021}. Approximating the hydrogen spin diffusion in the biphenyl crystal to that of a spin-1/2 species on a cubic lattice, $D_n \approx 0.22 \frac{\mu_0}{4\pi}\hbar \gamma_n^2\rho_n^{1/3}$ \cite{Cheung1981}, we find $D_n \approx 571$~nm$^2$/s. Thus diffusion out of the the NV sensing volume will happen on a comparable timescale to the polarisation transfer, leading to efficient polarisation of the total bath but only a small buildup of polarisation local to the NV. This is not a problem for the ultimate implementation of the technique but it does make unambiguously measuring polarisation buildup using the same shallow NVs difficult.

\begin{figure}
\includegraphics[width=0.4\textwidth]{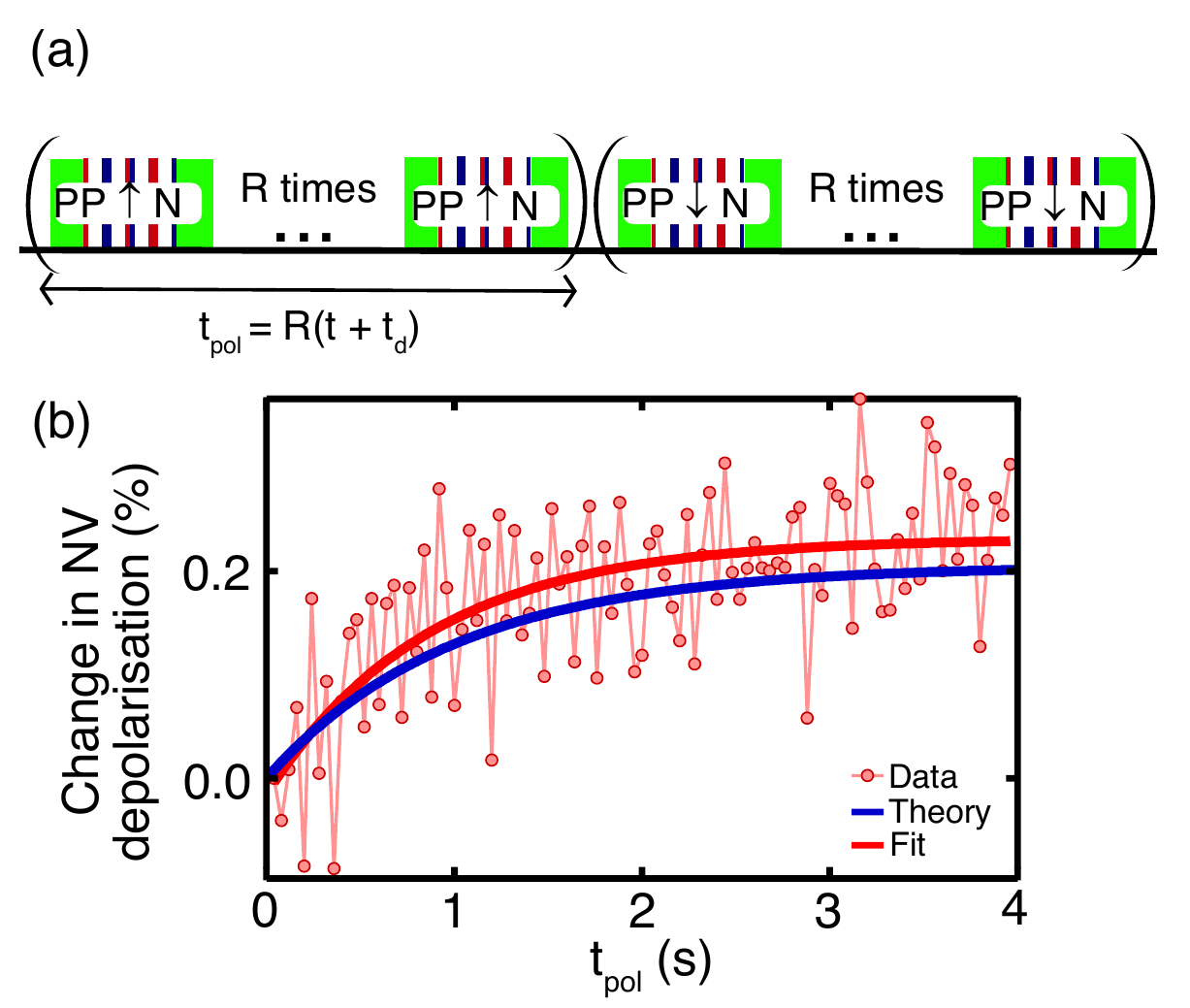}
\caption{\textbf{a)} Pulse sequence used to demonstrate local polarisation buildup: a resonant PulsePol sequence is repeated $R$ times to reach a steady state. The bath is then reset with a ``depolarisation" sequence of the same length, where polarisation in the opposite direction is achieved by initialising the NV in the $m_s=-1$ spin state prior to applying each PulsePol sequence. \textbf{b)} Measured polarisation buildup within NV sensing volume following repeated application of the PulsePol sequence previously identified to give the maximum cooling rate ($N=8$). The blue line is the theoretical prediction based on the solution of the differential equation Eqn. \ref{eqn: diffeq} with $T_{1,n}=1$~s and the red line is a fit to the experimental data, finding $T_{1,n}\approx 0.88$~s.}
\label{fig: buildup}
\end{figure}

Fig. \ref{fig: buildup} shows an attempt to evidence polarisation buildup within the NV sensing volume using the optimal sequence previously identified ($N=8$). This measurement is made as per the pulse sequence schematic in Fig. \ref{fig: buildup}(a): a resonant PulsePol sequence is repeated for a time $t_{\text{pol}}>T_{1,n}$ and polarisation buildup measured through the change in NV PL. The NV PL should increase over the course of the measurement as polarisation builds up within the NV sensing volume and the depth of the hydrogen resonance decreases (due to an effective reduction in $A_0$). The procedure is then repeated with the NV initialised in the $m_s = -1$ spin state to reset the nuclear bath.

 The solid blue curve in Fig \ref{fig: buildup}(b) is a theoretical prediction obtained by solving Eqn. \ref{eqn: diffeq} using the experimentally determined $u =7506$~s$^{-1}$ and setting $T_{1,n} = 1/\Gamma_{1,n} = 1$~s. $T_{1,n}$ sets the time scale to reach the steady state (here we use a simple exponential rise with time constant $T_{1,n}$). Higher values of $T_{1,n}$ will allow higher levels of polarisation to accumulate but this will result in only a modest increase in the steady state value as we make a local measurement and $D_n$ is large compared to $u$. Biphenyl $T_{1,n}$ has been reported to be as high as $\approx 10$~min at room temperature and $\approx 2$~T \cite{Liu1985}, but it is likely to be much shorter at our field ($\sim 450$~G).  
The steady state PulsePol dip reduction (the expected measurement contrast) for an NV 6~nm from the surface was calculated by numerically integrating over the steady state probability distribution. The predicted contrast change is extremely small ($\approx 0.2\%$) despite a significant overall polarisation of 1800 equivalent hydrogen spins per NV due to the role of spin diffusion in driving polarisation out of the NV sensing volume.

The experimental data (pink circles) are consistent with the theory but the signal is within noise. Each data point in Fig. \ref{fig: buildup}(b) is the difference in PL obtained between `signal' and `reference' camera exposures (that differ by a $\pi$ pulse prior to readout to account for common mode noise) and thus represents the cumulative PL given by $R \approx 2000$ repetitions of the PulsePol sequence. The signal to noise ratio is small as measurement times are necessarily long (hours to days), over which time additional noise from magnetic field and temperature shifts are difficult to control. Compounding this is the long laser pulse duration used (30~$\mu$s), necessary to ensure maximal and even ensemble initialisation across the full FOV but reducing the spin readout contrast. This also reduces the polarisation duty cycle in the experiment which is a major limiting factor in the steady-state polarisation produced, taking $u\mapsto u(t_{\text{seq}})\frac{t_{\text{seq}}}{t_{\text{seq}}+t_d} \cal F$, where $t_{\text{seq}}$ is the optimal sequence length and $t_d$ is the sequence ``dead time", exceeding 30~$\mu$s in this case, and a finite initialisation fidelity $\cal F$ $= 0.8$ has been included. A higher laser power density would address this issue and measuring polarisation buildup is thus expected to be easier on a confocal microscopy system, where the laser is focussed to a diffraction-limited spot. However, the lower power density in our experiment is necessary to access a wide field of view. Fortunately, this is a technical requirement that could be overcome in a realistic implementation by simply using an appropriate high-intensity pumping source and thus it is possible for $t_d$ to be negligible. The solid red curve in Fig. \ref{fig: buildup}(b) is a fit to the experimental data (again a single exponential rise), yielding a similar total contrast change on the slightly shorter time scale $T_{1,n} \approx 0.9$~s.

An alternative scheme to detect local polarisation involves pulsing the nuclear ensemble with a resonant $\pi /2$ pulse. This induces an AC field due to the spins' Larmor precession that is stronger than the statistical polarisation signal, which can then be read out using the NV AC magnetometry. This has the advantage of separating the polarisation and readout processes and may produce a measurable signal even though the local polarisation is not expected to be sufficient to produce a signal to noise ratio (SNR) enhancement compared to the continuous acquisition of statistical polarisation \cite{Tetienne2021}. However, this will still be a small signal with predicted strength similar to that read out using the polarisation sequence, so we do not expect that this would achieve a dramatically improved result. 
\subsection{Polarisation scaling with NV $T_2$}
\label{sec: nv t2}
Having demonstrated and analysed polarisation transfer in a current best case scenario, we now further probe the dependencies of Eqn. \ref{eqn: combined} by varying the NV and target properties. To see this clearly, we remove the influence of the background NV decay as much as possible by, for every data point, dividing the difference between the on- and off-resonance measurements by the off-resonance value at that point, to obtain a normalised polarisation signal $S(t)\equiv P_H(t)/P_{\rm off}(t)$. Taking our earlier definitions and Eqn. \ref{eqn: combined} we find
\begin{equation}
\label{eqn: normalised pol}
\begin{split}
S(t) = &1 - \exp(-\Gamma_{\rm int}t) \cos ^2 \left(\frac{A_0t}{2}\right) \\ &- \left(1-\exp (-\Gamma_{\rm int}t)\right)\exp \left(-\frac{A_0^2 t}{\Gamma_2^{\text{tot}}}\right),
\end{split}
\end{equation}
so that we have an exponential rise to unity with the possible appearance of sinusoidal behaviour at short times. Although we consistently observe saturation to a value below one due to the increased influence of decoherence at long times, this equation well describes our data for times $\leq 120~\mu$s which is well beyond a realistic optimal sequence length $\approx 10~\mu$s. 

\begin{figure}
\centering
\includegraphics[width=0.35\textwidth]{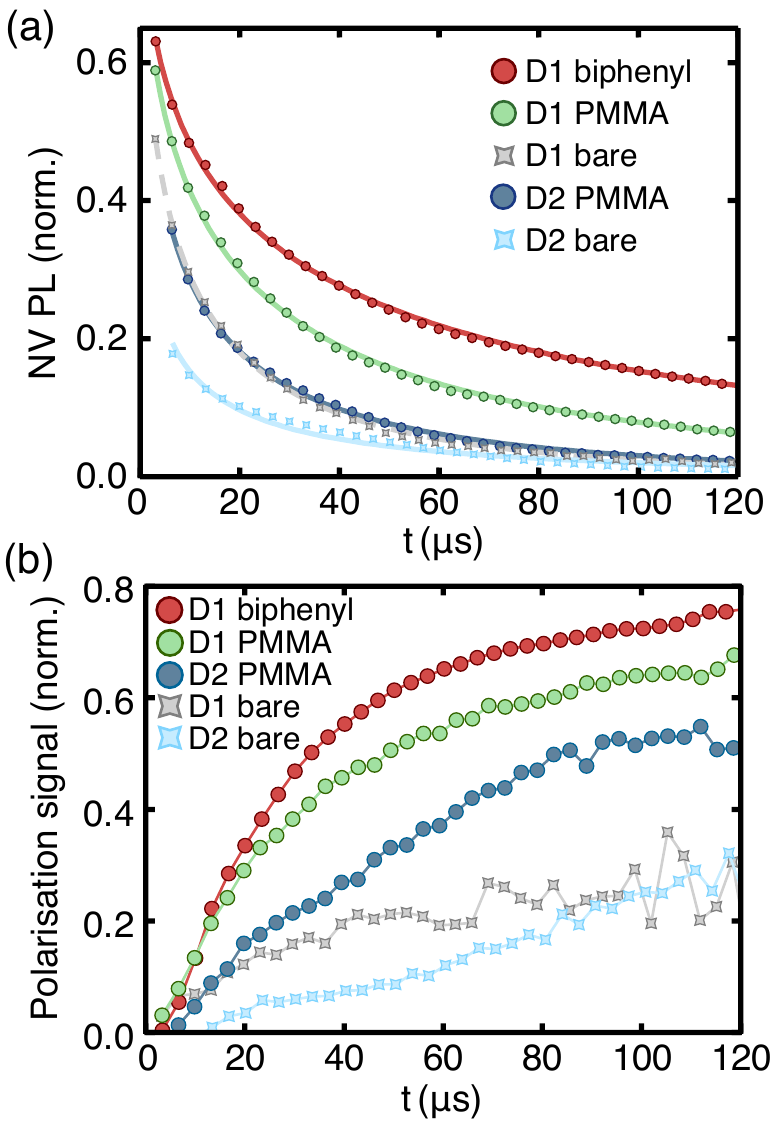}
\caption{\textbf{a)} Comparison of background (off-resonance) PulsePol $T_2$ decay for different solid targets (biphenyl, PMMA, bare diamond surface) and diamond samples (labelled D1 and D2). Solid lines are fits to an exponential decay $\exp (-(t/T_2^{NV})^{\beta})$ with $\beta$ ranging from 0.42 to 0.47. Data normalised to the maximum NV spin contrast for a given diamond. \textbf{b)} Normalised polarisation transfer for different solid targets and bare diamond hydrogen signals for two diamond samples.}
\label{fig: solid comparison}
\end{figure}
First we examine the influence of NV $T_2$, which still enters into Eqn. \ref{eqn: normalised pol} through the previous definition of $\Gamma_2^{\text{tot}}$. In Fig. \ref{fig: solid comparison} we present the normalised polarisation transfer signal obtained for two solid hydrogen targets, biphenyl and poly(methyl methacrylate) (PMMA), and using two diamond samples with different $T_2$ - varying between $\approx 7~\mu$s and 22~$\mu$s (with minimal variation in the exponent $\beta$) as shown in the off-resonance PulsePol decay curves in Fig. \ref{fig: solid comparison}(a). $T_2$ changes between two diamonds, D1 and D2, as D2 was implanted with twice the dose of $^{15}$N ions as D1 and so features a more dense nitrogen spin bath and possibly more implantation-induced crystal damage (see SI for diamond sample details). Additionally, however, $T_2$ was also found to vary based on the hydrogen target present on the diamond surface. This effect was confirmed to be a genuine background effect (rather than off-resonant polarisation transfer) by the appearance of similar variation in Hahn echo $T_2$ (see Appendix \ref{protocol comparison}) and has previously been observed in near-surface single NVs, where it was explained in terms of an electronic passivation of surface electron density \cite{Kim2015}. 

This shows that, even where the nitrogen spin bath is expected to be the dominant source of NV decoherence as in our samples, unpaired electron spins on the diamond surface (or other surface effects) are also significant. The dielectric properties of PMMA and biphenyl are unlikely to explain such a large difference but a similar electrical passivation may be possible through a separate mechanism such as chemisorption to unpaired electron spins on the surface through, for example, a target double bond \cite{Mamatkulov2006,Dubois2007}. Regardless, the source of this effect is not important for the current discussion, though it conveniently allows us to probe the dependence of the polarisation dynamics on NV $T_2$ in a controlled way.

The effect of the target-specific NV $T_2$ is evident through comparison of the D1 biphenyl and D1 PMMA curves in Fig. \ref{fig: solid comparison}(b). Polarisation transfer to biphenyl is observed to be more efficient than to PMMA, despite PMMA's greater hydrogen density ($\rho_H^{\text{PMMA}} = 56~$nm$^{-3}$ versus $\rho_H^{\text{biph.}} = 41~$nm$^{-3}$) and negligible differences in $T_{2,H}^*$ ($=1/\Gamma_2^H$ for a solid target) for both targets as measured via XY8 correlation spectroscopy \cite{Staudacher2015}. The difference in NV $T_2$ is sufficient to explain our data and is most evident at longer times when the system's evolution is well described by the strong-dephasing limit. For short times ($<10~\mu$s), when coherent evolution is prevalent, PMMA gives a stronger signal than biphenyl as the coherent polarisation exchange is related only to the coupling strength $A_0 \propto \sqrt{\rho_H}$. 

Comparing the PMMA signal from the two diamonds, it is clear that a reduction in NV $T_2$ reduces per-NV polarisation transfer efficiency (possibly small differences in coupling strength too due to depth distribution - see Appendix \ref{characterisation}). This results in a 3.5-fold reduction in maximal experimentally measured per-NV cooling rate. This is greater than the difference in estimated NV density between the two samples, although there is room to improve the properties of high-density NV ensembles (see Sec. \ref{sec: discussion}). A similar difference is also observed in the respective bare diamond signals, which is believed to arise from an adventitous hydrogen layer on the diamond surface \cite{DeVience2015a,Staudacher2015}. This signal is expected to be present in all our data as it is not removed by any cleaning process. We do not subtract it from the data in this work as in principle any polarisation transferred to this layer can diffuse into the target and, regardless, its contribution to the maximum cooling rates for solid targets is small.
\subsection{Effect of target diffusion}
\label{sec: liquid}
The results of Sec. \ref{sec: nv t2} demonstrate the impact of changing $\Gamma_2^{\text{tot}}$ through varying NV $T_2$. Differences in the target $T_2$ will affect polarisation transfer in the same way. While the measured $T_{2,H}^*$ values for the two solid targets compared were similar, molecular diffusion of target spins into and out of the NV sensing volume during the polarisation transfer will result in a loss of phase coherence in the interaction (and thus an effective $T_2$, scaling as $D_n^{-1}$). When using a pulse-based scheme such as PulsePol this is compounded by an additional loss in polarisation transfer efficiency, as the requirement for the NV-target interaction to be long enough lived for the average Hamiltonian to accurately approximate a flip-flop Hamiltonian introduces an effective reduction in coupling which scales as $D_n^{-2}$ when the diffusion timescale exceeds the Larmor period \cite{Hall2020}. We therefore expect the cooling rate $u_{\rm max} \propto A_0^2  /\Gamma_2^{\rm tot}$ using PulsePol to scale as $D_n^{-3}$ for fluid targets.

This represents a challenge for genuine liquid-state hyperpolarisation, which is an attractive niche for NV-based techniques to fill as current liquid hyperpolarisation techniques are less efficient than their solid-state counterparts \cite{Berthault2020}. Additionally, NV-based micron-scale NMR has recently emerged as having utility for analysis of liquid samples \cite{Glenn2018,Smits2019}, and has recently been successfully integrated with Overhauser DNP and parahydrogen-based hyerpolarisation techniques \cite{Bucher2020,Arunkumar2021}. It has been proposed that an all-diamond hyperpolarisation and NMR platform could be competitive in this regime \cite{Tetienne2021}. The goal of this section, then, is to experimentally determine the efficiency of polarisation transfer to targets in the liquid state compared to those in the solid state, with diamond material properties held constant.

To assess the impact of molecular diffusion on polarisation transfer, several mixtures of glycerol and water were produced with the glycerol volume fraction $\chi_g$ ranging from 0 (pure water) to 1 (pure glycerol). Assuming the static coupling between bath hydrogen spins and NV is constant (justifiable as $\rho_H$ should vary by less than 1\% under laboratory conditions \cite{Volk2018}) and any effect of changing background NV decoherence due to target dielectric properties is minimal (see Appendix \ref{diffusion} : NV $T_2$ changes minimally), any loss in polarisation transfer efficiency with reduced viscosity can be ascribed to the changing molecular dynamics. 
\begin{figure*}
\includegraphics[width=0.8\textwidth]{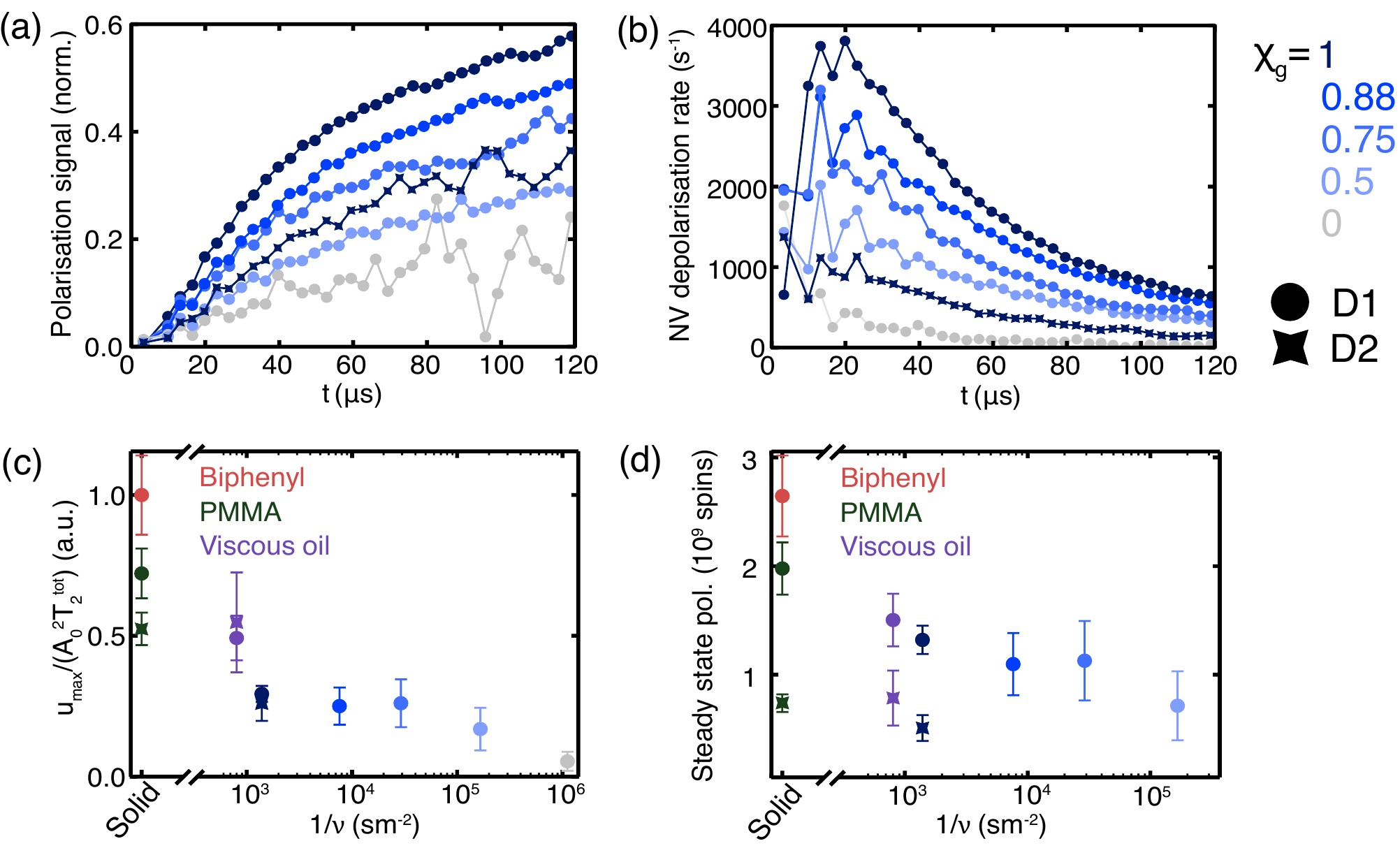}
\caption{\textbf{a)} NV polarisation transfer to glycerol-water mixtures of various compositions denoted by glycerol volume fraction $\chi_g$, normalised to background decoherence. \textbf{b)} Experimental cooling rates from the data in a). \textbf{c)} Plot of maximum cooling rate, normalised with respect to the theoretical no-diffusion value, against the inverse of the kinematic viscosities of the glycerol mixtures and additional liquid and solid state targets for comparison. Error bars extrapolated from the range of cooling rates obtained within $N=N_{\text{opt}}\pm 2$ where $N_{\text{opt}}$ corresponds to the maximum cooling rate. \textbf{d)} Steady-state nuclear polarisation values predicted by a simulation using the measured cooling rates, NV densities, and the addressed field of view of 50~$\times$~50~$\mu$m. Idealised parameters were used, including $\cal F$ = 0.8, $t_d = 0~\mu$s, and $T_{1,n}=1$~s.}
\label{fig: fluids}
\end{figure*}

Fig. \ref{fig: fluids}(a) plots the inferred polarisation transfer, with the background NV decay (diamond D1) normalised out for clarity. A clear trend is visible, with the growth of the pure glycerol resonance comparable to a solid target due to its high viscosity, while diluting with water down to $\chi_g = 0.5$ gives reduced but still measurable polarisation transfer. Using pure water as a target, on the other hand, gives no measurable increase in signal over the bare diamond hydrogen resonance. We also include a measurement obtained using diamond D2 and a pure glycerol target for comparison with the effect of changing NV $T_2$, which is consistent with the results of the previous section. 

Fig. \ref{fig: fluids}(b) plots the NV depolarisation rates (which we again take to correspond to the cooling rate) as in Sec. \ref{sec: biphenyl}. Compared to the biphenyl results, the maximum cooling rate for pure glycerol is roughly a factor of two lower, despite almost identical background NV $T_2$. XY8 correlation spectroscopy measures only a small reduction in target $T_2$ which is not enough to explain the full reduction, especially when considering that glycerol is more hydrogen-dense than biphenyl ($\rho_H^{\text{gly}}\approx 65$~nm$^{-3}$). Thus our data supports an additional reduction in polarisation transfer as a result of using the PulsePol sequence: assuming again that the maximum experimental cooling rate is proportional to the theoretical strong-dephasing limited rate, this result suggests a reduction in effective coupling strength by a factor of $\approx 3$. However, Fig. \ref{fig: fluids}(b) shows that reducing NV $T_2$ by using D2 instead of D1 made a larger difference than diluting glycerol by 50\% by volume in water, which is expected to result in an increase in $D_n$ of over an order of magnitude \cite{DErrico2004}.

Fig. \ref{fig: fluids}(c) summarises the cooling rate results for the glycerol mixtures and other targets for comparison. As well as the solid targets already discussed, we also include results obtained with a highly viscous oil (Sigma-Aldrich 1250~cSt I0890 immersion oil). As a figure of merit, we divide the maximum measured cooling rate by the theoretical strong-dephasing cooling rate for each data point, normalising to the maximum value obtained (biphenyl). In doing so, we normalise out any differences due to $T_2^{\text{tot}}$, $d_{NV}$, or $\rho_H$, isolating the effective coupling reduction due to molecular diffusion \cite{Hall2020}. Again we include data obtained with diamond D2 as well as D1 for three of the targets, finding that differences between the diamonds are well normalised out, providing further support for the model. 

This quantity is plotted against the inverse of the kinematic viscosity, a readily available quantity that is proportional to $D_n$ through the Stokes-Einstein equation (although this proportionality will not necessarily be identical for all targets). The remaining trend confirms that increasing molecular diffusion makes polarisation transfer to liquid targets less efficient than to solid targets using PulsePol. However, the dependence in the intermediate viscosity regime (glycerol series: $\chi_g = 0.5-1$) is weaker than expected. 

This is also true for the variation of measured $T_2^H$ ($=1/\Gamma_2^H$ for a fluid target) for the series of glycerol mixtures, where a reduction is present but with weaker scaling than the expected $D_n^{-1}$. Within error, our data is consistent with either variation in $D_n$ of only approximately 20\% between $\chi_g = 1$ and $\chi_g=0.5$, or with the full series exhibiting characteristic diffusion times less than the hydrogen Larmor period ($\approx 550$~ns) (see Appendix \ref{diffusion}). In both cases this contradicts expectation \cite{DErrico2004}. However, the discrepancy may be explained either in terms of a near-surface adsorption layer where fluid motion is not well described by bulk viscosity values, or possible near-surface solidification, both of which have previously been suggested \cite{Staudacher2015,Kim2015}.


Fig. \ref{fig: fluids}(d) plots the total steady-state nuclear polarisation predicted to be produced by the maximum experimentally determined cooling rates and taking into account the measured $\sigma_{NV}$ for the respective samples ($\sigma_{NV} = 1400~\mu$m$^{-2}$ for D1 and 1800~$\mu$m$^{-2}$ for D2) and the area addressed (50~$\times$~50~$\mu$m). As diffusion drives polarisation away from the NV sensing volume as in Sec. \ref{sec: biphenyl} in all cases, local saturation effects are negligible and the simulated steady state hydrogen polarisation per NV tends towards $u(t_{\text{seq}})\frac{t_{\text{seq}}}{t_{\text{seq}}+t_d} \cal F$ with $T_{1,n}=1$~s. Here we plot the limiting case of $t_d=0$. As this data is not normalised against $A_0^2/\Gamma_2^{\text{tot}}$, the effect of NV $T_2$ can again be seen here, with the slight increase in NV density in D2 insufficient to offset the reduction in $T_2$ in this sample. 

These results confirm the difficulty of achieving polarisation to highly fluid targets, with no evidence found for significant interaction with water. However, in the intermediate viscosity region, the observed dependence is weaker than theoretically predicted. Sufficiently viscous liquids diffuse slowly enough to interact with strength within a factor of two of that of solid targets and so in this regime NV based hyerpolarisation appears to hold promise. This could be applicable to micron-scale NV NMR modalities and hyperpolarisation of more fluid targets could be achievable by exploiting the variation of viscosity with temperature; polarising a slow-moving liquid at low temperature and then heating up, similar to a freeze-thaw cycle commonly used for solution-state DNP but without the need for a convenient phase transition. Additionally, the apparent reduction in molecular diffusion near the diamond surface could be exploited to achieve greater polarisation transfer than would be expected for a fluid with true bulk properties. Although this effect requires more rigorous characterisation, it is likely that it is confined to within a few nm of the diamond surface and so molecular and/or spin diffusion will quickly take polarisation into the bulk fluid. 

\section{Discussion}
\label{sec: discussion}
Our results show that the coherence properties in current dense, shallow NV ensembles are sufficient to transfer a significant amount of polarisation to an external nuclear spin bath, in both solids and sufficiently viscous liquids. The PulsePol protocol is successful in allowing the use of $\approx 10^6$ NVs in parallel and the extension of the current experiment to an even larger active area is feasible provided technical challenges of microwave delivery and high-intensity laser illumination are met over that area. A higher intensity source for the NV optical pumping could allow a significant decrease in the laser pulse duration used, resulting in an improved polarisation duty cycle that would provide an improvement in the steady-state hydrogen polarisation over the current experiment. 

In the limiting case of achieving perfect NV initialisation instantly, the cooling rate demonstrated with biphenyl is sufficient to generate an average polarisation of $P=2.7\times 10^{-4}$ within the 1-$\mu$m-thick layer closest to the diamond (taking the uniform NV density $\sigma_{NV}=1400~\mu$m$^{-2}$ and $T_{1,n}=1$~s). This represents an over three orders of magnitude enhancement over the Boltzmann polarisation under our laboratory conditions, owing mainly to the low magnetic field used ($\sim$~450~G) and operation at room temperature. The absolute polarisation is small, however, and would not scale with magnetic field or temperature where a brute-force Boltzmann polarisation enhancement could be obtained. Furthermore, practical realisation of this technique for bulk NMR enhancement would require structuring a diamond polarisation cell such that the volume above the diamond slab is of a micron scale or smaller, which represents a significant engineering challenge \cite{Tetienne2021}. However, there are also significant material improvements that can be made to enhance the polarisation obtained. 

The NV yield in both diamond samples used in this study (1.4\% for D1 and 0.9\% for D2) is low compared to that of typical deeper ensembles \cite{Healey2020} and it has been suggested that techniques such as Fermi engineering could enhance this yield towards 100\% \cite{Luhmann2019}. In the latter case, an improvement of nearly two orders of magnitude may be achieveable in the area-normalised polarisation generated, whereas even in the former case a factor of 2-5 improvement is possible. The difference between yields in shallow and deeper ensembles can be understood in terms of band bending due to imperfections in the oxygen-terminated diamond surface. Improving this is an active area of research, for example there are current efforts to understand and control the diamond surface through optimised cleaning and annealing strategies \cite{Sangtawesin2019}.

Reducing near-surface band bending will also result in a shallower mean NV depth, resulting in stronger coupling with the external spin bath and larger cooling rates (scaling with $d_{NV}^{-3/2}$ in the coherent case and $d_{NV}^{-3}$ in the strong-dephasing limit). The dependence shown in Sec. \ref{sec: nv t2} of NV $T_2$ on hydrogen target also shows that, even with the high nitrogen densities of the samples used, surface spins contribute significantly to NV decoherence. The combined action of reducing the mean ensemble depth to that of the expected ion implantation range of a 2.5~keV implant $\approx$4~nm and increasing NV $T_2$ to a nitrogen-limited value would, according to our theoretical model outlined in Sec. \ref{sec: biphenyl}, result in an additional cooling rate enhancement of upwards of four times. 

Control over the diamond surface also has a final benefit: unpaired surface electron spins also contribute to the dephasing of the hydrogen spins closest to the diamond \cite{Hall2020}. We included this factor in the the analysis of Sec. \ref{sec: biphenyl}, taking a typical surface electron density of $\sigma_e=0.1$~nm$^{-2}$ \cite{Rosskopf2014,Stacey2019} which reproduced our experimental results well. Numerical integration of the quantity $A(\textbf{R})^2/\Gamma_2^{\text{tot}}(\textbf{R})$, the cooling rate in the strong-dephasing limit, shows that this electron density results in a 44\% drop in the effective coupling between an NV 6~nm from the surface and the hydrogen bath (for biphenyl). 

Considering all of the above factors, we estimate that a total improvement of up to two orders of magnitude is achievable with realistic diamond material advancements. This would be sufficient, following successful diamond nanostructuring and integration with an NMR probe, to achieve significant enhancements over thermal polarisation \cite{Tetienne2021}.  

The results of Sec. \ref{sec: liquid} confirm that NV hyperpolarisation of low-viscosity liquid targets is beyond current capabilities but the cooling rates obtained for more viscous targets were within a factor 2-3 of the solids. The transfer appears to be aided by an effective reduction in molecular diffusion near the diamond surface, possibly due to surface dragging effects or local drying brought about by laser heating. This surprising result may be beneficial for polarisation to fluids with moderate viscosity ($D_n\approx 10^{-11}$~m$^2$s$^{-1}$), whereby polarisation can be transferred to regions of locally low diffusion and allowed to diffuse into the bulk fluid. Diffusion constants of this order of magnitude are typical in liquid crystal and lipid bilayer targets \cite{Tauber2013,Gaede2003}, which could be appealing targets for NV-based polarisation. In particular, if polarisation transfer to these targets remains within an order of magnitude of solid targets as in our experiment, the improvements to the diamond materials discussed above could provide a realistic avenue towards hyperpolarised NV-based micron-scale NMR \cite{Tetienne2021}. Further work is required to fully characterise the near-surface fluid dynamics, however. We emphasise that we do not have a direct measurement of the diffusion constants of the targets used in our experiments (local or bulk) and so the interpretation of our current results can only be qualitative.
\section{Conclusion}
We have presented the first reported experimental study of external hyperpolarisation using an NV ensemble. We demonstrated evidence of polarisation transfer in the closest to ideal case, to hydrogen nuclei within a solid target, finding agreement between experiment and an analytical model combining coherent and incoherent transfer mechanisms. We found that the polarisation rate obtainable with current materials and with realistic experimental conditions is sufficient to achieve a modest (three orders of magnitude over room temperature, low field Boltzmann conditions) enhancement over thermal polarisation levels although spin diffusion prohibits direct detection within the NV sensing volume. Still, this result is compatible with the vision of achieving meaningful NMR enhancement with a shallow NV ensemble following realistic diamond material improvements. 

Polarisation of nuclear targets within the liquid state proved to be more challenging due to the theoretically predicted reduction in effective dipolar coupling strength inherent to this technique in the presence of molecular diffusion. However, although polarisation of highly diffusive liquids (e.g. water) is unrealistic, our results show polarisation transfer efficiency to more viscous fluids within an order of magnitude of that to solid targets. This is a promising result that could mean NV-based liquid state hyperpolarisation is viable following successful incorporation of a diamond hyperpolariser with a liquid NMR probe. Other methods of polarisation transfer such as lab frame cross relaxation are expected to perform better than PulsePol in the fluid target regime \cite{Hall2020}, although this protocol faces other challenges such as poor initialisation fidelity $\cal F$ at the resonance condition \cite{Broadway2016a}. 

Considering the above points, the single largest driver of improved future prospects for NV-based NMR enhancement will be centred around improving diamond material properties, particularly regarding the diamond surface. Our characterisation of the diamond samples used in this study suggests that near perfect control of diamond surface chemistry could result in multiple orders of magnitude improvement in polarisation transfer efficiency using the PulsePol sequence, as well as the possibility of using other protocols. 

\section*{Acknowledgements}

We acknowledge support from the Australian Research Council (ARC) through grants DE170100129, CE170100012, and DP190101506. A.J.H. and G.A.L.W. are supported by an Australian Government Research Training Program Scholarship. T.T. acknowledges the support of JSPS KAKENHI (No. 20H02187 and 20H05661), JST CREST (JPMJCR1773) and MEXT Q-LEAP (JPMXS0118068379).
\appendix
\section{Experimental details}
\label{expt detail}
All experiments were carried out on a purpose-built widefield NV microscope, elements of which are depicted schematically in Fig. \ref{schematic} of the main text. A 532~nm laser with an incident intensity of $\approx 300$~mW was used to excite and initialise the spin states of the NV ensembles. Red photoluminescence (PL) is emitted in a spin state-dependent manner by the NVs as they relax back to the ground state and imaged onto a scientific complementary metal oxide semiconductor (sCMOS) camera. This setup allows for imaging of the NVs' behaviour under a variety of experimental protocols over a maximum of 200~$\times$~200~$\mu$m field of view with a diffraction-limited spatial resolution of $\approx 400$~nm. 

Quantum sensing and polarisation-inducing protocols typically require the delivery of radio frequency (RF) radiation to the NV ensemble in order to manipulate its spin state. This is achieved in our setup via a gold ring-shaped resonator deposited on a glass coverslip onto which the diamond sample is mounted. The RF radiation is delivered using a Rhode \& Schwartz SMBV100A signal generator IQ modulated by a Keysight P9336A arbitrary waveform generator (AWG) with 1~ns time resolution. A PulseBlaster ESR-pro card controls the timing of RF pulse sequences, laser pulses, and camera triggers with a time resolution of 2~ns. 

The diamond samples used in this study are electronic grade chemical vapour deposition (CVD) substrates purchased from Delaware Diamond Knives. A 1~$\mu$m thick layer of isotopically enriched (99.95\%) $^{12}$C diamond was overgrown via microwave plasma-assisted CVD \cite{Teraji2015} to both limit spin noise and eliminate spurious harmonic signals due to $^{13}$C spins in NV NMR experiments \cite{Loretz2015}. The shallow NV ensembles used for sensing and polarisation were then created via a low energy $^{15}$N ion implantation procedure (2.5~keV, dose 1-2$\times 10^{13}$~cm$^{-2}$, InnovIon) and annealed using a ramp sequence culminating at 1100 $\degree$C to maximise NV yield and ensemble coherence properties \cite{Tetienne2018}. Oxygen surface termination was achieved using a boiling mixtures of sulfuric and nitric acid. The result is a high density NV ensemble with mean distance from the diamond surface of $d_{\text{NV}} \approx 7$~nm (see Appendix \ref{characterisation}).
\section{Sample and target characterisation}
\label{characterisation}
Two diamond samples were used for this study, one featuring a 2.5~keV $^{15}$N implant with a dose of 1$\times 10^{13}$~cm$^{-2}$ (sample D1) and the other a $^{15}$N implant of the same energy but higher dose of 2$\times 10^{13}$~cm$^{-2}$ (sample D2). The NV$^-$ density of these samples was estimated by comparing their fluorescence as measured on a confocal microscope to that given by a single NV (in a separate sample) and assuming the measured PL is purely from the negative charge state. The estimated overall N to NV$^-$ conversion ratios in these samples were 1.4\% (D1) and 0.9\% (D2), lower than typical conversion rates in deeper ensembles which can exceed 5\% \cite{Healey2020}. 
\begin{figure*}[htb]
\centering
\includegraphics[width=0.9\textwidth]{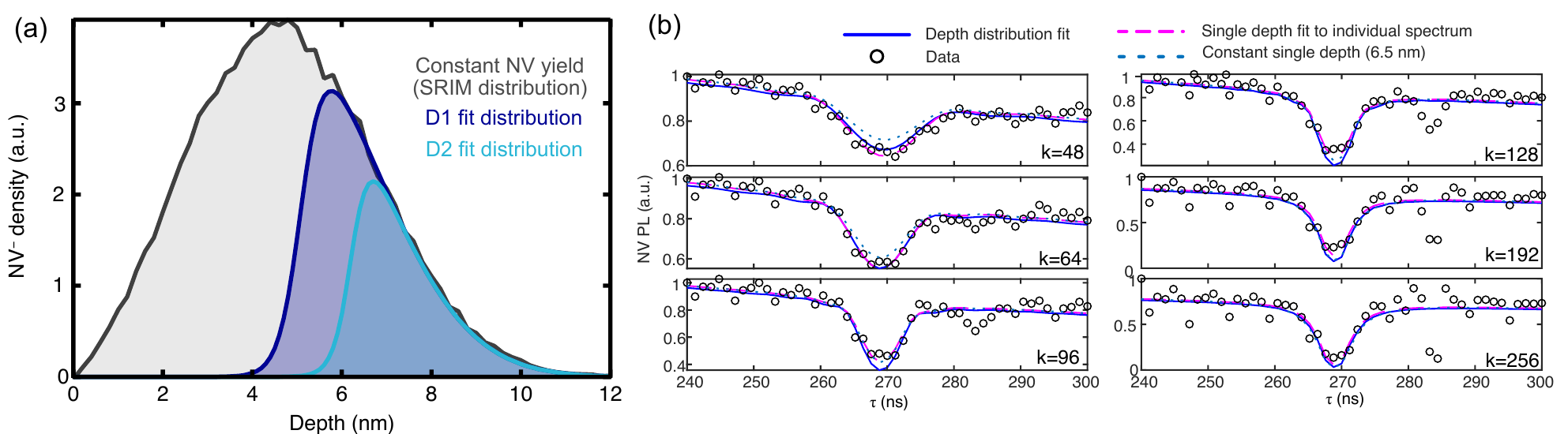}
\caption{\textbf{a)} Comparison of fit NV$^-$ depth distributions for samples D1 (blue) and D2 (cyan) to simulated N implant distribution (grey, same for both samples up to a factor of two). \textbf{b)} Set of XY8-$k$ spectra used to fit the depth distribution for sample D1, where $k$ is the total number of $\pi$ pulses used in the sequence. The signal reconstructed from the fit distribution (solid blue lines) gives good but not perfect agreement with the experimental data, indicating that a larger data set would be required to perfectly reconstruct the NV depth distribution. Fitting directly to individual spectra using a single discrete depth as in \cite{Pham2016} (dashed pink curves) results in a better fit to the data but a variable NV depth depending on $k$. Using a constant discrete depth (dashed light blue curves), chosen to match the mean of the fit distribution ($\approx 6.5$~nm), provides reasonable agreement with experiment except at low $k$. This indicates that while knowledge of the full distribution is required to successfully reproduce the signal for all sequences, using the mean depth captures the main behaviour, justifying the analytical approach used in the main text.} 
\label{sifig: depthdist}
\end{figure*}

Following the method of \cite{Ziem2019}, we measure the depth distribution of the NV ensembles in our samples. We fit the pairwise differences of 5-6 XY8-$k$ spectra with $k$ ranging from 32 to 256 to the signal given by a model NV distribution. The NV distribution was modelled as the product of a Gaussian fit to the expected N depth distribution as simulated using the stopping range in matter (SRIM) software package (mean 4.6~nm, width 3.0~nm) and a Sigmoid function $\left(1+\exp (-p_1 (x-p_2))\right)^{-1}$ representing an effective cut-off in NV$^-$ charge stability due to band bending. The slope $p_1$ and position $p_2$ of the cut-off were allowed to vary in the fitting routine, as was a final parameter $p_0$ which multiplies the distribution to ensure normalisation. The distributions obtained are likely not uniquely defined by only a small number of XY8 spectra and we only use approximate mean depths in our analysis in the main text. Using the simpler method \cite{Pham2016} and obtaining a single mean depth is consistent with our approach provided $k > 64$ and so would be sufficient for this work. However, it is useful to know that our NV distribution is well described by a known implanted N distribution and band bending, and motivates further work in improving near-surface NV yield through diamond surface engineering. The reduced overall NV yield is largely explained by this cut-off, suggesting that the local NV yield could be constant for depths greater than $\approx 7$~nm depending on the sample. NV charge stability is known to depend on the local electron donor density as well as the Fermi level position, however, and it is also possible that N:NV conversion is reduced close to the surface due to the surface acting as an efficient vacancy trap during annealing, so the task of optimising near-surface NV production is not trivial. 

Example distributions are shown in Fig. \ref{sifig: depthdist} alongside the implanted nitrogen distribution predicted by a SRIM simulation. The fit distributions are consistent with there being a constant NV$^-$ yield (likely similar to that of deeper ensembles) for depths deeper than some sharp cut-off that is dependent on the surface properties of individual samples. For shallower depths than this cut-off, the NV$^0$ charge state is expected to dominate. Here we see that sample D2 has a deeper cut-off than D1, matching the difference in conversion ratios estimated from the samples' PL in comparison to that given by a stable single NV$^-$ centre. The difference could be due to the higher implantation dose used on D2 creating greater damage to the diamond surface and explains why less polarisation transfer was measured with this sample (even accounting for the greater N density).
\begin{figure}
\includegraphics[width=0.3\textwidth]{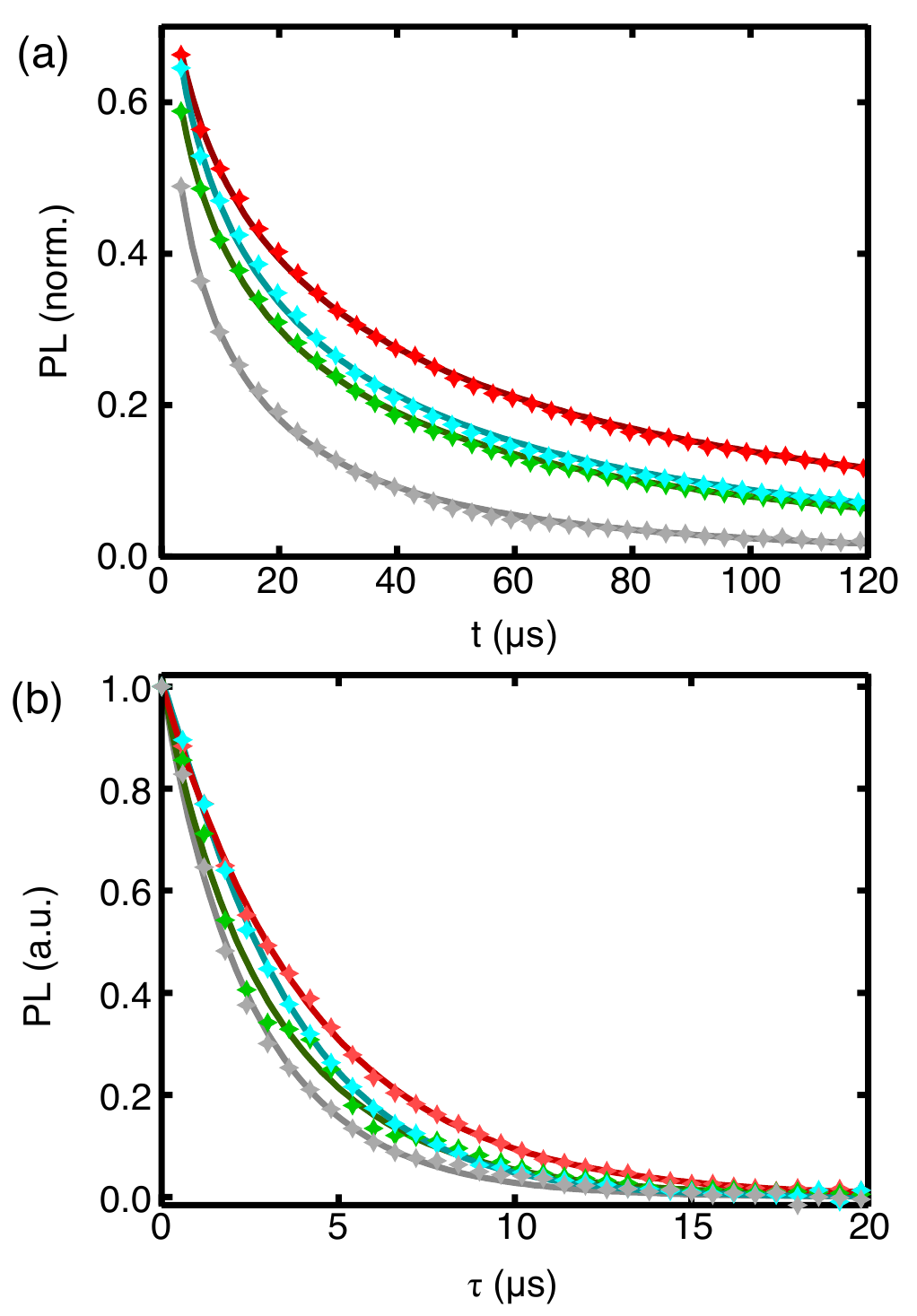}
\caption{\textbf{a)} Off-resonance PulsePol decays with glycerol (red), viscous oil (blue), and PMMA (green) on the diamond surface compared to in air (grey). Solid lines are fits to the equation $\exp \left( -(t/T_2)^{\beta}\right)$ with $\beta \approx 0.47$ in each case. Data normalised to the maxiumum NV spin contrast. \textbf{b)} Hahn echo decays for the same targets, showing a similar $T_2$ trend. $\tau$ defined as the total length of the sequence. Solid lines are fits to single exponential decays.}
\label{sifig: targett2}
\end{figure} 

Most liquid hydrogen targets were simply deposited onto the surface of the diamond where they were observed to be stable for the duration of the experiments. A polydimethylsiloxane (PDMS) well and shorter total measurement times were used for the water experiment to prevent evaporation. PMMA was deposited on the diamond and then allowed to cure at room temperature prior to measurement. Biphenyl crystals were deposited by dissolving biphenyl powder (Sigma-Aldrich, 99.5\% purity) in isopropanol and allowing this solution to dry on the diamond surface. To prevent degradation of the crystals, they were encased within water and UV-curing epoxy. The resulting crystals, visible in bright-field images such as Fig. 1(b), were observed to be stable for the duration of the experiments. Between measurements of different targets, diamonds were cleaned first using an appropriate solvent (dichloromethane, acetone, or ethanol) and then in a boiling mixture of sulfuric and nitric acid to ensure a clean, oxygen-terminated surface. In each case the hydrogen signal was observed to return to the bare diamond baseline.

\begin{figure*}[htb]
\includegraphics[width=0.9\textwidth]{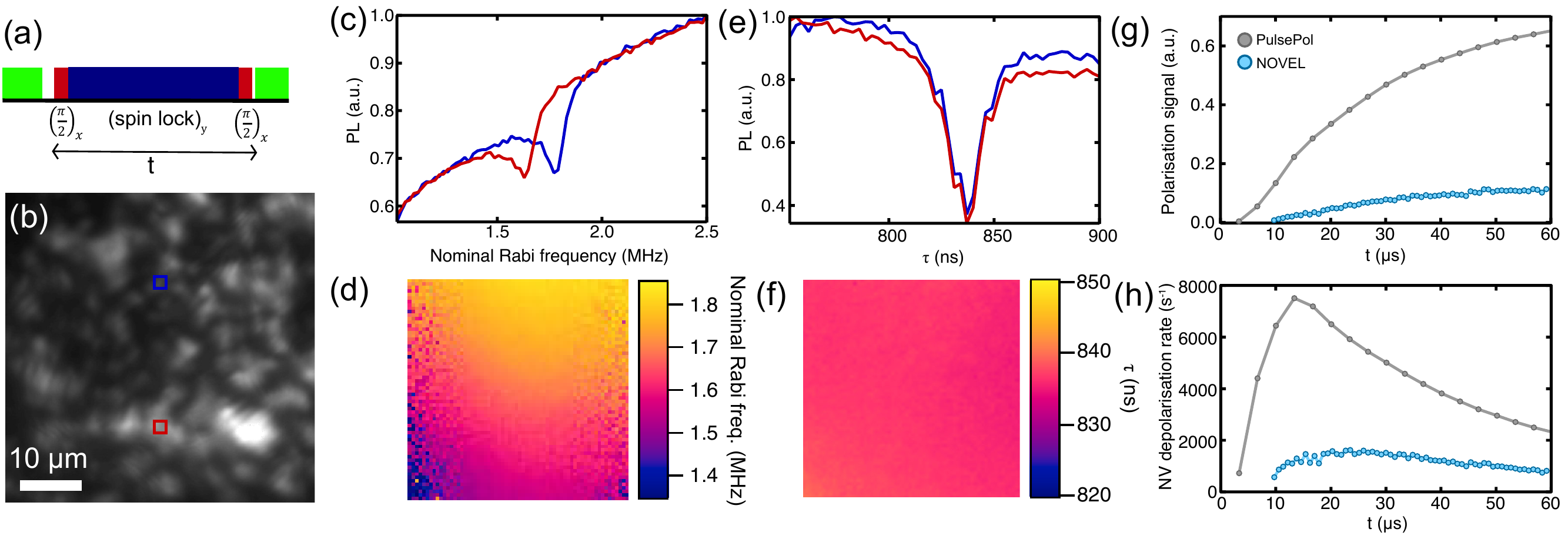}
\caption{\textbf{a)} NOVEL pulse sequence schematic. \textbf{b)} NV PL image of 50 $\times$ 50 $\mu$m FOV with areas for comparison highlighted by blue and red squares. \textbf{c)} Spectra obtained using NOVEL with spin locking time $t = 20$~$\mu$s showing hydrogen resonances (biphenyl). Red and blue traces are data obtained over the regions highlighted by red and blue squares in (b) respectively. \textbf{d)} Colourmap of full 50~$\times$~50~$\mu$m FOV showing position of hydrogen resonance at each pixel, given by fitting to a Lorentzian dip shape. \textbf{e)} Same as c) but for a $\tau$ sweep using the PulsePol sequence as defined in the main text ($N=30$). \textbf{f)} Colourmap of PulsePol hydrogen resonance position, fit as in d). \textbf{g)} Polarisation signal normalised to background NV decoherence  obtained using PulsePol (grey) and NOVEL (light blue). NOVEL data taken over a $\approx 2~\times ~2~\mu$m region while PulsePol data is averaged over the full FOV. \textbf{h)} Cooling rates obtained (using the same definition as in the main text) using PulsePol (grey) and NOVEL (light blue).} 
\label{sifig: novel}
\end{figure*}

The correlation time $\tau_c$ of the target spin bath can be measured using XY8 correlation spectroscopy \cite{Staudacher2015}. For the hydrogen spins in the solid samples, $\tau_c$ directly gives the nuclear dephasing time $T_{2,n}^*$ as molecular motion can be neglected. For the fluid targets the opposite is true: $\tau_c$ is governed primarily by spatial diffusion. These measured parameters are used in the theoretical calculations in the main text.

Fig. \ref{sifig: targett2} shows the variation of NV $T_2$ with hydrogen target obtained using both off-resonance PulsePol sequences (a) and the Hahn echo sequence (b). The $T_2$ scaling is similar in both cases, showing that the variation is due to a changing noise spectrum rather than polarisation behaviour. 

\section{Protocol comparison}
\label{protocol comparison}
As discussed in the main text, the PulsePol protocol was selected for this work due to its suitability in addressing an NV ensemble over a wide field of view. To illustrate this, we compared the behaviour of PulsePol and the well known NOVEL sequence \cite{Henstra2008}, depicted in Fig. \ref{sifig: novel}(a), over an identical region (Fig. \ref{sifig: novel}(b)) with biphenyl as the hydrogen target. Panels (c) and (e) show the appearance of the hydrogen resonance using the NOVEL and PulsePol sequences respectively, with red and blue traces showing data averaged over the pixels contained within the corresponding regions marked on the NV PL image (b). The NOVEL spectra are obtained by using a spin locking time of 20~$\mu$s and sweeping the microwave power, which corresponds to a Rabi frequency in the rotating frame. The hydrogen resonance appears when, locally, the Rabi frequency matches the hydrogen Larmor frequency and indicates the transfer of spin polarisation from NV to the hydrogen bath. A microwave power gradient across our FOV results in the resonance appearing for different nominal driving powers between the two regions. Fitting the position of the resonance for every pixel in the image, shown in panel (d), confirms that there is a sharp gradient in resonance position across the whole FOV, meaning that is is impossible to transfer polarisation using more than 1\% of this region at any one time under our experimental conditions. 

On the contrary, the spectra in Fig. \ref{sifig: novel}(e) overlap, and the colourplot Fig. \ref{sifig: novel}(f) confirms that the PulsePol resonance position does not vary by more than its linewidth across the entire FOV, owing to the sequence's robustness to microwave power detuning. The PulsePol resonances are also much deeper, corresponding to more efficient polarisation transfer, despite the theoretical effective coupling being reduced by 28\% compared to NOVEL. This is due to the fact that PulsePol achieves $T_2 >20~\mu$s whereas resonant $T_{1\rho} \approx 2~\mu$s with $B\sim 450$~G. It is also possible that even the power gradient over a single $\mu$m scale is enough to reduce the efficiency of transfer using NOVEL. 

Fig. \ref{sifig: novel}(g) and (h) show the difference between off- and on-resonance NV decay, normalised to the background decoherence and converted to a cooling rate respectively, for the two sequences (PulsePol shown in grey and NOVEL in light blue). As in the main text, the PulsePol signal is averaged over the entire FOV while the NOVEL signal is averaged over only the blue region. The influence of the longer PulsePol $T_2$ is clear, with much more efficient polarisation transfer obtained per NV with PulsePol than with NOVEL. The maximum cooling rate per NV obtained using NOVEL is 1604 spins/sec, 79\% less than that obtained using PulsePol. Again assuming a proportionality between these values and the theoretical strong-dephasing cooling rate and recalling that $A_0(\text{PulsePol}) = 0.72 A_0(\text{NOVEL})$ suggests $T_{1\rho} < T_2^{PP} /10$, in agreement with the measured values.

Considering that NOVEL is only active over $\approx 1\%$ of the FOV, the steady-state hydrogen polarisation possible to be enacted using this sequence in our experiment is likely to be approximately three orders of magnitude lower than that obtained using PulsePol.

\section{Diffusion data}
\label{diffusion}
\begin{figure}
\includegraphics[width=0.3\textwidth]{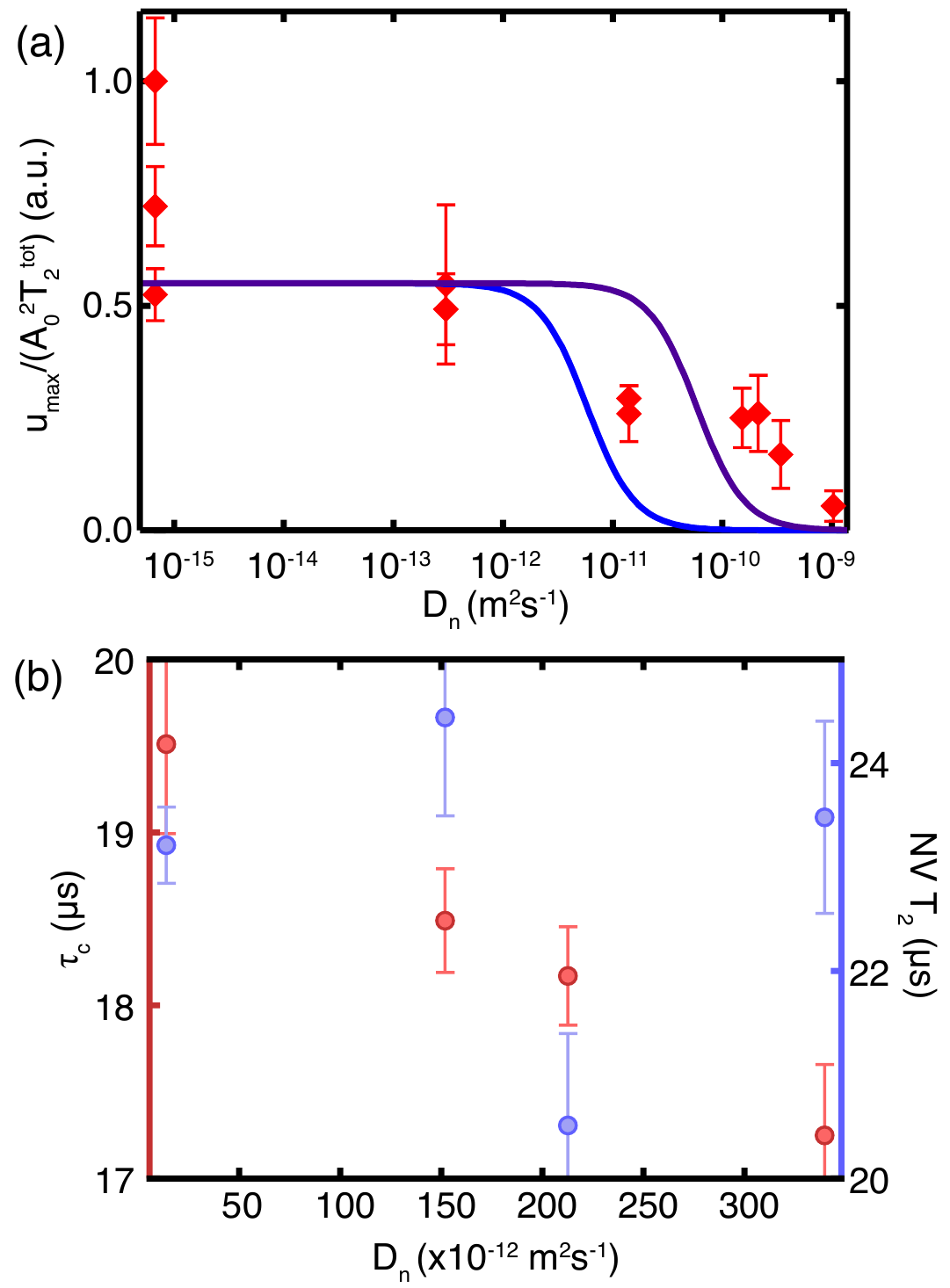}
\caption{\textbf{a)} Normalised cooling rate plotted against diffusion constant $D_n$. Solid lines show a theoretical prediction for NV-H distances of 7~nm (blue) and 14~nm (purple). \textbf{b)}Variation of correlation time $\tau_c$ measured by XY8-64 correlation spectroscopy (red points, left axis), and NV PulsePol $T_2$ (blue points, right axis) for glycerol/water mixtures with $\chi_g = 1,0.88,0.75,0.5$. Error bars show the standard fitting error. Diffusion constant $D_n$ given by equation in Ref. \cite{DErrico2004}}. 
\label{sifig: difft2}
\end{figure}
In Sec. \ref{sec: liquid} we presented data showing the reduction in polarisation transfer efficiency associated with increasing levels of molecular diffusion. Due to the uncertainty and potential inconsistency in reporting diffusion constants $D_n$ for different targets, we presented that data only in terms of the kinematic viscosity (readily available in the literature for glycerol mixtures and provided by the manufacturer for viscous oil). For completeness we plot the same data as in the main text against estimated values of $D_n$ in Fig. \ref{sifig: difft2}(a). For the glycerol mixtures, we take values from the empirically derived formula in Ref. \cite{DErrico2004}, while in the absence of manufacturer specification of molecular mass or hydrodynamic radius for the viscous oil, we use a value $D_n=3\times 10^{-13}$~m$^2$s$^{-1}$ estimated in previous work for a similar viscous oil \cite{Staudacher2015}. In the blue and purple solid lines we plot the curve $\frac{\tau_D^2}{\tau_D^2 + \tau_L^2}$ for NV-H distances 7~nm and 14~nm respectively, representing the approximate bounds of the NV sensing volume. This equation gives the expected reduction in effective coupling to the hydrogen spin bath as a result of using the PulsePol sequence \cite{Hall2020}, which should be the only remaining diffusion-dependent factor following the normalisation described in the main text. The maximum value was set arbitrarily, to match the solid target providing the least signal for illustration. 

We can see some qualitative agreement between experiment and theory, with the viscous oil diffusing slowly enough to not be greatly affected, while a relatively rapid drop-off in effective coupling is observed for $\chi_g\leq 0.75$. However, overall the points belonging to the glycerol/water mixture series are not well described by the theory. As mentioned in the main text, two possible adjustments to these data points could recover the expected trend: either a systematic decrease in $D_n$ of around one order of magnitude due to interactions between the diamond surface and the fluid, or less dramatic increase in diffusivity near the surface with decreasing $\chi_g$. The latter case could arise due to the local properties of the mixtures changing near surface, for instance due to laser heating causing water to evaporate leaving $\chi_g >0.9$ within the NV sensing volume in all cases. 

Support for both interpretations is found in experimental measurements of the bath correlation time via XY8-64 correlation spectroscopy \cite{Staudacher2015}, presented in Fig. \ref{sifig: difft2}(b) (red points, left axis), where we show only data for the glycerol mixture series (excluding pure water which gives a very poor signal). $\tau_c$ varies inversely with $D_n$ as expected, but not as sharply as predicted. Assuming $T_2^* \gg T_2^{\text{diff}} \approx \tau_c$, we expect $D_n^{-1}$ scaling, implying that $D$ varies by less than 20\% between $\chi_g = 1$ and $\chi_g = 0.5$. This would be consistent with a water evaporation effect, increasing the near-surface $\chi_g$. Additionally, taking the expression $\tau_c = 2d_{NV}^2/D_n$ \cite{Pham2016} and $\tau_c = 19.51~\mu$s obtained for pure glycerol gives $D_n=3.69\times 10^{-12}$~m$^2$s$^{-1}$ (for $d_{NV} = 6$~nm), well below the literature value.

\bibliographystyle{apsrev} 
\bibliography{hypbib}
\end{document}